\documentclass[
]{ceurart}

\usepackage{listings}
\usepackage{url}

\usepackage{caption}
\usepackage{subcaption}
\begin{document}

\copyrightyear{2024}
\copyrightclause{Copyright for this paper by its authors.
  Use permitted under Creative Commons License Attribution 4.0
  International (CC BY 4.0).}

\conference{IberLEF 2024, September 2024, Valladolid, Spain}

\title{BSC-UPC at EmoSPeech-IberLEF2024: Attention Pooling for Emotion Recognition}

\tnotemark[1]

\author[1]{Marc Casals-Salvador}[%
orcid=0009-0003-9099-3826,
email=marc.casals@bsc.es,
url=https://github.com/marccasals98,
]
\cormark[1]

\address[1]{Barcelona Supercomputing Center,  Eusebi Güell Square 1-3, 08034 Barcelona, Spain}

\author[2]{Federico Costa}[%
orcid=0000-0002-1389-3595,
email=federico.costa@upc.edu,
]

\author[2]{Miquel India}[%
orcid=0000-0002-3107-3662,
email=miquel.angel.india@upc.edu,
]

\address[2]{Universitat Politècnica de Catalunya,
  Jordi Girona 31, 08034 Barcelona, Spain}

\author[1,2]{Javier Hernando}[%
orcid=0000-0002-1730-8154,
email=javier.hernando@upc.edu,
]

\cortext[1]{Corresponding author.}

\begin{abstract}
The domain of speech emotion recognition (SER) has persistently been a frontier within the landscape of machine learning. It is an active field that has been revolutionized in the last few decades and whose implementations are remarkable in multiple applications that could affect daily life. Consequently, the Iberian Languages Evaluation Forum (IberLEF) of 2024 held a competitive challenge to leverage the SER results with a Spanish corpus. This paper presents the approach followed with the goal of participating in this competition. The main architecture consists of different pre-trained speech and text models to extract features from both modalities, utilizing an attention pooling mechanism. The proposed system has achieved the first position in the challenge with an 86.69\% in Macro F1-Score.

\end{abstract}


\maketitle

\section{Introduction}

Emotions are undoubtedly fundamental parts of our idiosyncrasy. They play an important role in interpersonal relationships and decision-making and generally take part in the evolution and consciousness of any mental process \cite{izard_emotion_2009}. Moreover, there is empirical evidence that emotions immensely influence human health \cite{consedine_role_2007}, which creates the necessity to monitor them. Affective computing is of great interest in medical health fields \cite{jaques_predicting_2017}. Consequently, developing a system capable of discerning various emotions is highly valuable. Researchers have attempted to predict emotions using machine learning approaches, but their effectiveness depends on the quality and quantity of available data.

In the field of Natural Language Processing (NLP), numerous emotion recognition models rely exclusively on text-based features.  Since the creation of Transformers \cite{vaswani_attention_2023}, multiple approaches \cite{adoma_comparative_2020, kumar_bert_2022, qin_bert-erc_2023}
 seek to use pre-trained Transformer encoders, such as BERT \cite{devlin_bert_2019} as text feature extractors. These encoders are pre-trained models using self-supervised learning techniques on extensive datasets. This pre-training enables the models to project words in a latent space with a rich semantic representation, from which classifiers can then be employed to predict the corresponding emotion classes. On the other hand, speech is crucial for expressing emotions. Elements such as pitch, prominence, and phrasing contribute generously to providing emotion information. As explained in \cite{pell_recognizing_2009}, the human brain is capable of recognizing emotions pan-culturally and independently of the language they are expressed with. Following this principle, the researchers have proposed different architectures to process and extract information from speech signals. In the realm of Speech Emotion Recognition (SER), Machine Learning (ML) and Deep Learning (DL) models often utilize hand-crafted features such as  Mel-Frequency Cepstral Coefficients (MFCC) \cite{ingale_speech_2012, xu_head_2021, singh_speech_2023, mohan_speech_2023}. Nevertheless, recent advances in Deep Learning allow cutting-edge architectures to combine text and speech to provide better results. Multimodal emotion Recognition (MER) is a complex task since it requires models to be able to learn complex patterns of the data. For this reason, the usage of text and audio pre-trained models significantly improves the embedding representation of their features, allowing them to be combined in their latent space \cite{macary_use_2021,pepino_emotion_2021, zhao_multi-level_2022}.

In summary, the exceptional progress that has been made in this field is appreciable, yet there also exists a disparity in the amount of work carried out in Spanish. Developing these models using Spanish data is crucial for several reasons. On the one hand, Spanish is one of the languages with the most native speakers worldwide. On the other hand, it is necessary to leverage the engineering opportunities of Spanish-speaking countries, opening the door to developing new technologies that could satisfy their population needs. However, one of the difficulties this presents is the lack of labelled data in Spanish needed to train any supervised learning approach.  

In order to encourage the creation of an emotion recognition model trained with Spanish data, the Iberian Languages Evaluation Forum (IberLEF) of 2024 created the challenge EmoSPeech 2024. This competition evaluates the Macro F1-Score of the participants in two tasks: Emotion Recognition with text and with speech and text. 
The training corpus is Spanish MEACorpus 2023 \cite{pan2024spanish}. This dataset contains 13.16 h of speech, and its transparent methodology distinguishes it from other datasets of the same task. The speech samples are collected from YouTube videos and are labelled using the categorical taxonomy proposed by P. Ekman \cite{ekman_facial_1979}, which include emotions such as surprise, disgust, anger, joy, sadness, fear, and neutral expressions. Nevertheless, it was impossible for the annotators to find any sample that contained speech expressing surprise emotion, though this class is not represented in the dataset.

This paper endeavours to leverage the research of Spanish models with cutting-edge technologies in the current state of the art by making use of the MEACorpus 2023. The current state of the art is the usage of Transformers-based pre-trained models with high capacity that, leveraging the enormous datasets they are trained with, are capable of describing complex patterns in both text and speech. Following this lead, the system combines a speech pre-trained model, the XLSR-wav2vec 2.0 \cite{conneau_unsupervised_2020}, that is trained with 436,000h and a RoBERTA text model fine-tuned in Spanish \cite{DBLP:journals/corr/abs-2109-08597}. Both models are used as feature extractors for speech and text respectively, and they output a vector with the relevant information of the utterances. The two vectors are then concatenated into a single vector that contains information about text and speech. Subsequently, this vector is reduced to lower-dimension representation by making use of an attention pooling mechanism. Finally, dense layers are utilized to project this reduced vector, determining the classification of the utterance. This approach has reached an F1-Score of 86.69\%, achieving the first position in the multimodal task of the EmoSPeech 2024 challenge.

\section{Challenge}

The EmoSPeech 2024 \cite{overviewEmoSPeech2024} is a Challenge proposed by the Iberian Languages Evaluation Forum (IberLEF) of 2024 \cite{overviewIberLEF2024}, which is a Spanish workshop hosted by the \textit{Sociedad Española para el Procesamiento del Lenguaje Natural} (SEPLN). This event is dedicated to producing models in the frame of the Iberian peninsula, encompassing different national languages such as Spanish, Portuguese, Catalan, Basque, or Galician. 

To develop the competition, the organisers proposed a dataset named Spanish MEACorpus 2023 \cite{pan2024spanish}. The dataset, comprising 13.16h of speech divided into 5,129 audios, was meticulously labelled by the research team of the article. As explained in their paper, the procedure to extract the audio files is as follows: The authors of the dataset selected YouTube videos according to their topic and extracted audio segments considering the noise of the audio files and the silence gaps. Once this part was done, the audio files were classified using the emotion taxonomy developed by P. Ekman \cite{ekman_facial_1979}. It comprehended the basic emotions of disgust, anger, joy, sadness, fear, surprise, and neutral emotion. Nevertheless, despite the efforts, finding any speech audio that contained the surprise emotion was impossible.  

\begin{figure}[h]
    \centering
    \includegraphics[width=\textwidth]{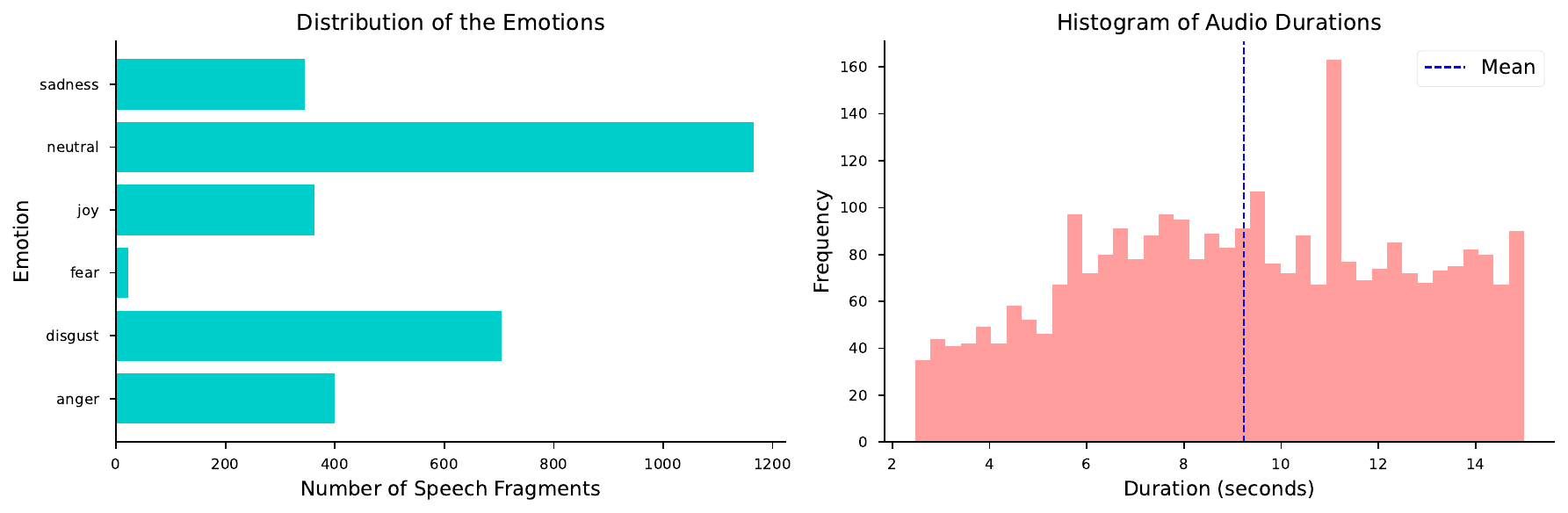}
    \caption{Some of the characteristics of the MEACorpus 2023. On the left is the distribution of the number of speech fragments over emotions. On the right, a histogram of the durations of the audio fragments.}
    \label{fig:emotion-distribution}
\end{figure}

As is common in the field, the dataset exhibits an unequal representation of the emotion classes. Figure \ref{fig:emotion-distribution} (left)shows that neutrality and disgust are the most prevalent emotions, while fear is notably scarce. Another important aspect is the length of the audio files, which can directly affect the performance of the network by providing more contextual information about the speech. Therefore, audio duration is an essential characteristic of any speech dataset and must be considered a quality metric. A histogram of the audio file duration is shown in Figure \ref{fig:emotion-distribution}(right). The mean of the duration is 9.24 s. Lastly, another fundamental consideration is the variability of the recordings. The fact that this competition uses third-party audio files makes it more difficult to control other parameters, such as noise or the magnitude of the audio. Some videos are recorded outdoors and are more likely to have background noise or exhibit poorer recording quality, while others are studio-recorded and of higher quality. Furthermore, the dataset's paper detailed that 46\% of the speech segments are attributed to female voices, with the remainder belonging to males. The paper affirmed that the text transcriptions were extracted directly from the raw audio file using Whisper \cite{radford2022robust} followed by a manual revision by the researchers.

\section{Architecture}
\label{sec:architecture}

\begin{figure}[h]
    \centering
    \includegraphics[width=0.4\textwidth]{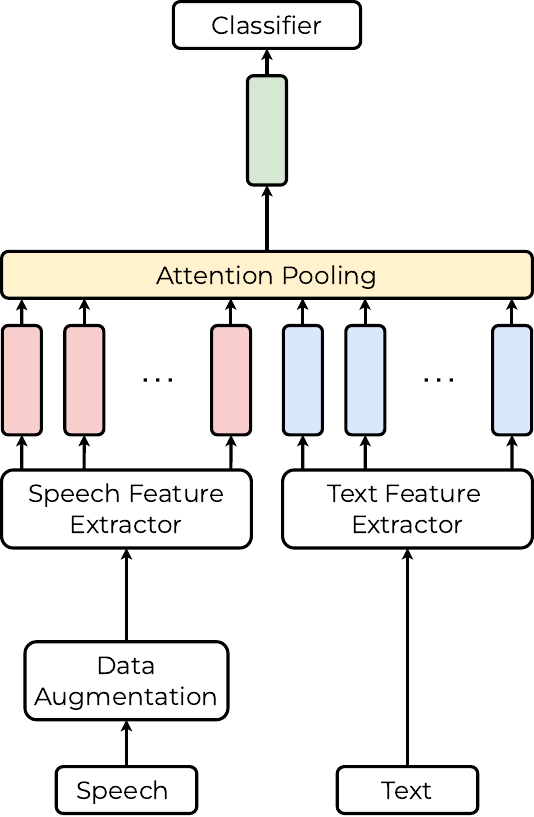}
    \vspace{0.5cm}    
    \caption{Diagram Attention Pooling for the Multimodal Emotion Recognition System. The speech utterances are represented in red and the text is represented in blue.}
    \label{fig:system-diagram}
\end{figure}

The system created is a multimodality model that combines text and speech and, trained with the Spanish MEACorpus 2023 dataset \cite{pan2024spanish}, effectuates a classification of the emotion. In Figure \ref{fig:system-diagram}, it is possible to see a global representation of the system. As can be seen, both text and speech are fed to the network and processed with a self-supervised learning (SSL) model that works as a feature extractor. Then, the model concatenates these features and pools them into one single vector. Finally, a classification is performed by processing this vector with multilayer perceptrons that serve as a classifier.

\subsection{Speech}
It is a standard practice to apply regularisation techniques to preprocess the data before its utilisation in a machine learning model. In the case of this system, the overall mean and standard deviation of all the audio files were calculated and used to normalise the values in the dataset. This technique is widespread in Deep Learning literature for speech due to the fact that it makes backpropagation more efficient and reduces the impact of the outliers. The normalisation process was extended to the validation and test sets, whereby the mean and standard deviation values obtained from the training set were utilised. It is essential to mention that, during the training stage, the samples were randomly cropped with a window of 5.5 seconds. The shorter utterances were enlarged using repetition padding. 

Given the limited size of the dataset, it became necessary to use data augmentation to mitigate overfitting to the training set. The specific techniques utilized included speed perturbation, which alters the speed of the audio; reverberation, which simulates a reverberant environment; and background noise, which adds ambient sounds to the audio. PyTorch implements data augmentation using the following scheme: first of all, defining the transformations used to create synthetic data. Then, for each sample, a probability decides whether the model will see the original data or the one created after applying these transformations. This process occurs in every epoch, so data augmentation is applied to different samples in each epoch. This streaming data augmentation guarantees the model is exposed to a wide diversity of data without physically increasing the dataset size. As is common in the field, when doing inference is not desired to perturb the data, so in the validation and test sets, the probability of applying these transformations is zero.  

Recent advancements in the field of Deep Learning have highlighted the importance of using Transformer-based pre-trained models. It has been proven that their ability to adapt to changes in the domain makes them suitable for extracting features from audio files in any dataset. In this work, the following models were experimented with:

\begin{itemize}
            \item \textbf{WavLM} \cite{chen_wavlm_2022}: This cutting-edge model is trained with 80,000 hours and encompasses datasets such as Libri-Light \cite{Kahn_2020}, GigaSpeech  \cite{chen2021gigaspeech} and VoxPopuli \cite{wang2021voxpopuli}. Two versions of this model were tried, the Large version and the Base. The output vector is 768 in the case of the base version and 1,024 in the case of the large. 
    
    \item \textbf{XLSR-wav2vec 2.0} \cite{conneau_unsupervised_2020}: This model is based on wav2vec 2.0  \cite{baevski_wav2vec_2020} and it is trained with the datasets Common Voice \cite{ardila2020common}, BABEL  \cite{Gales2014SpeechRA} and Multilingual LibriSpeech \cite{pratap_mls_2020}, which makes a total of 436,000 hours of audio in 128 languages. This model outputs a vector of dimension 1,024.

    \item \textbf{HuBERT} \cite{hsu_hubert_2021}: This model is trained with 60,000 hours of Libri-Light \cite{Kahn_2020}. The output vector is of dimension 1,024.
\end{itemize}

\subsection{Text}
The text domain was the first to employ pre-trained large language models (LLMs) for different tasks. Since the creation of the BERT model \cite{devlin2019bert}, different approaches have emerged in the state of the art, following the same idea with some variations. In the approach of this work, the following SSL models were experimented with:
\begin{itemize}
    \item \textbf{BERT} \cite{devlin2019bert}: The large uncased version was used, with an an output dimension of 1,024.
    
    \item \textbf{XLM-RoBERTa Spanish}
    \cite{lange-etal-2021-meddoprof}:
    It is a pre-trained model based on XLM-RoBERTa \cite{conneau2020unsupervised} and trained with Spanish Unannotated Corpora  \cite{jose_canete_2019_3247731}. This model outputs a vector with 1,024 dimensions.

    \item \textbf{BETO} \cite{CaneteCFP2020}: BETO is one of the first pre-trained models produced with Spanish data. It follows the same structure as the BERT base. Consequently, it outputs a hidden vector of 768 dimensions. It is trained using Wikipedia data and all of the sources of the OPUS Project \cite{tiedemann-2012-parallel}. Additionally, a fine-tuned version of BETO for emotions was used in the system.
\end{itemize}

\subsection{Classifier}

Following the attention pooling process, the resultant vector undergoes further processing via a classifier module. This module is designed to include a layer that adjusts the vector's dimension to fit the desired hidden layer width. It also consists of a stack of hidden layers and an output layer that has a dimension equivalent to the number of classes being considered.

The model architecture consists of several linear layers, each followed by a dropout, layer normalization, and a Gaussian Error Linear Unit (GELU) activation function \cite{DBLP:journals/corr/HendrycksG16}. The Softmax activation function is used in the output layer to select the predicted class.

\section{Attention Pooling}

Different approaches have emerged for integrating information extracted from pre-trained models in the field of multimodal learning. Attention mechanisms, particularly Multi-Head Attention (MHA), have been popular in recent years for combining text and speech utterances. This study used an alternative Attention Pooling mechanism used in works such as \cite{DBLP:journals/corr/abs-1906-09890, costa2024double, india2021double} to reduce the dimensionality of the hidden state vector created by concatenating the outputs of the two pre-trained models. 
 
\begin{figure}[h]
    \centering
    \includegraphics[width=0.9\textwidth]{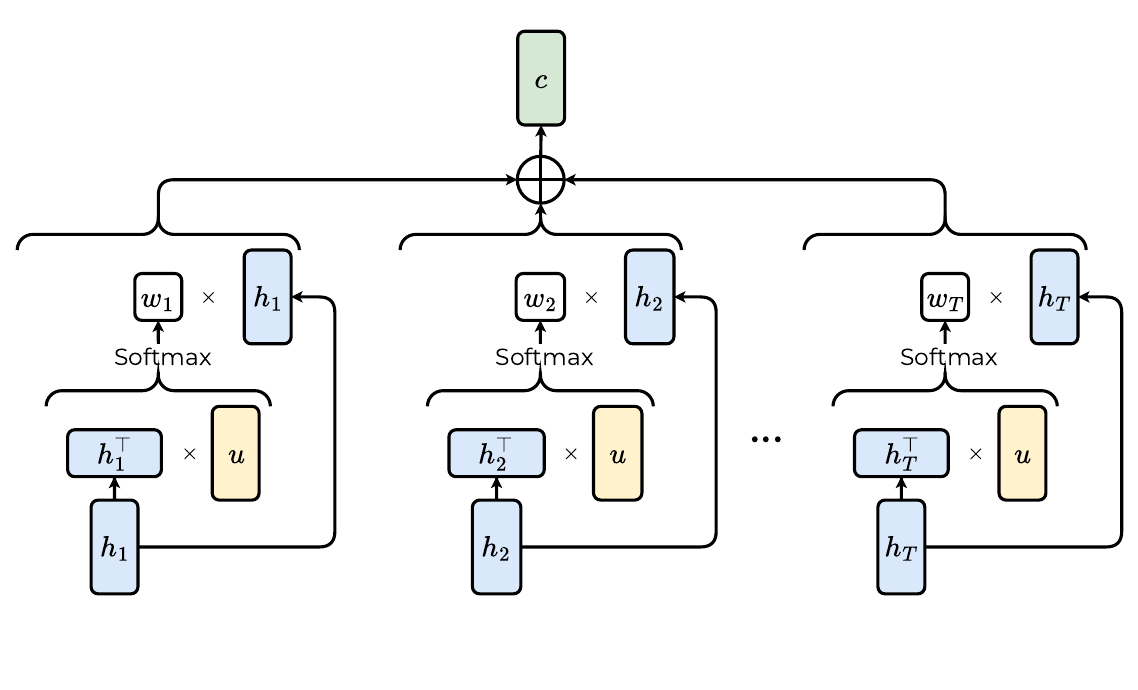}
    \caption{Diagram of the Attention Pooling operation. The hidden vector $h=h_1,...,h_T$ is pooled into a lower dimension $c$, using a learnable parameter $u$.}
    \label{fig:attention-pooling}
\end{figure}

Considering the embedding dimension $E$ and a batch size of one, we define the hidden states as the sequences of the extracted features $\{h_t\in\mathbb{R}^E|t=1,...,T\}$. Then, for each hidden state $h_t$ we calculate its weight as described in Equation \eqref{eq:weight-calculation}:

\begin{equation}
    w_t = \frac{\exp\left(\cfrac{h_t^\top u}{\sqrt{E}}\right)}{\sum_{i=1}^T\exp\left(\cfrac{h_i^\top u}{\sqrt{E}}\right)}
    \label{eq:weight-calculation}
\end{equation}

where $u\in\mathbb{R}^E$ is a trainable parameter initialized with the Xavier initialization \cite{pmlr-v9-glorot10a} and $w_t$ is the weight associated at the hidden vector $h_t$. Then, the pooled representation of the hidden vector is calculated using Equation \eqref{eq:weighted-sum}.

\begin{equation}
    c = \sum_{t=1}^{T} w_t h_t
    \label{eq:weighted-sum}
\end{equation}

The vector $c$ encapsulates the relevant information of the features extracted in the text and speech systems. This approach is computationally more efficient compared to the general Attention mechanism, where the key, query, and values are calculated.  This characteristic is especially convenient for this system due to the scarce data provided. Figure \ref{fig:attention-pooling} demonstrates graphically the functioning of the attention pooling mechanism. 

\section{Experimental Setup}
\label{sec:experimental-setup}

PyTorch requires all samples in the batch to have the same dimensions. Therefore, during training, the audio files were cropped using a window of 5.5 seconds, which was the optimal value found. In inference, the whole audio waveform was used. As detailed in Section \ref{sec:architecture}, the audio waveforms were normalized using the mean and standard deviation of the training set. The values extracted were -33.62 and 56.15, respectively. These same values are applied to the other sets when doing inference. Data augmentation techniques were applied by varying the probability based on the capacity of the model. The optimal value was found to be 0.3.

A batch size of 16 samples was selected, as it provided an optimal trade-off between minimizing the duration of each epoch and avoiding GPU memory exhaustion. To further accelerate computation, data parallelization across two GPUs was utilized. In particular, the GPUs employed were two NVIDIA GeForce RTX 2080Ti. The optimizer selected was the  AdamW \cite{loshchilov2019decoupled}, with a learning rate of 0.00005, which decayed by 10\% after five epochs without improvements in the validation F1-score. The dropout rate, set at 0.1, was adjusted according to the network's complexity. The number of epochs utilized also depended on the model's capacity. Although model parameters were only stored when the F1-score improved, early stopping was necessary due to the noisy and variable learning curves. This variability could lead to an overfitting model being saved based on local improvements in the F1-Score during validation. Each experiment lasted one to two days, depending on the configurations.

To improve our position on the leaderboard, we made Macro F1-Score our top priority since it was the metric used to evaluate the participants' submissions. This metric combines the Precision and the Recall into one single number by applying their harmonic mean. Specifically, the Macro F1-Score is the average of the F1-Score of each class. This metric treats all classes equally, regardless of their amount of data, making it a fair measure of overall performance. Given that the metric of interest is the macro F1-Score, it is imperative to mitigate the class imbalance present in this dataset. A wide variety of losses try to palliate this disparity. Beyond these possibilities, the loss criterion finally chosen is the weighted cross-entropy loss.

After doing the hyperparameter search, two classical machine learning techniques were employed with the aim of improving the results. The first approach involved applying thresholds to modify the final decision over the logits. However, this did not enhance the results. The second strategy leveraged the variability of different models by using a 3-model ensemble. Hard voting was chosen as the ensemble technique, where the most voted prediction among the three models was selected. In the event of a tie, the prediction from the model with the highest F1-Score on the validation set was chosen. The code of the project is available here: \url{https://github.com/marccasals98/BSC-UPC_EmoSPeech}

\section{Results}
\label{sec:results}
 In the initial stages of the competition, it was necessary to evaluate various pre-trained self-supervised models to determine the most suitable one for the data. At this moment, only the training corpus was available; therefore, it was necessary to make a validation partition to evaluate the performances of the different self-supervised models. Table \ref{tab:results-feature-extractors} presents the best results obtained with the diverse text and feature extractors. The best configuration used RoBERTa for text and XLSR-wav2vec 2.0 for audio, achieving an F1-score of 89.73\% on the validation set. This superior performance is likely because RoBERTa was trained on Spanish data, making it more effective for this domain than other models trained in English. In addition, XLS-wav2vec 2.0 was trained by using 436,000 hours of audio in 128 languages, including Spanish, which possibly contributed to the improvement of this metric score. It is worth noting that, despite BETO being a Spanish version of BERT, the results with this encoder were poor. This fact could be attributed to WavLM not being trained with as much Spanish data as XLSR-wav2vec 2.0  or that BETO's vector dimension is 768 instead of 1,024, resulting in fewer features captured by the model.

\begin{table}[h]
\begin{tabular}{llll}
\toprule
\textbf{Text Model} & \textbf{Audio Model} & \textbf{Output Dimensions} & \textbf{Validation F1-Score} \\
\hline
RoBERTa             & WavLM LARGE          & 1,024                       & 80.04\%                       \\
\textbf{RoBERTa}             & \textbf{XLSR-wav2vec 2.0}& \textbf{1,024}                       & \textbf{89.73\%}                       \\
RoBERTa             & HuBERT LARGE         & 1,024                       & 76.033\%                       \\
BERT Large Uncased  & WavLM LARGE          & 1,024                       & 83.27\%                       \\
BERT Large Uncased  & XLSR-wav2vec 2.0 & 1,024                       & 86.59\%                       \\
BETO                & WavLM BASE PLUS      & 768                        & 74.79\%                       \\
BETO-EMO            & WavLM BASE PLUS      & 768                        & 73.19\%\\
\bottomrule
\end{tabular}
\caption{Different results were obtained during the validation test with the different feature extractor models. All of these configurations had their corresponding hyperparameter tuning, and the best of each one was selected.}
\label{tab:results-feature-extractors}
\end{table}

It is remarkable that, in this initial stage, some experiments were conducted with other architectures that involved more parameters. For instance, attempts were made to combine features extracted from the encoder models using multi-head attention (MHA) with one and two heads. These experiments yielded unsatisfactory results, with F1-scores of 84.4\% and 82.2\%, and the models exhibited significant overfitting. Consequently, it was decided to discontinue these lines of experimentation and concentrate all efforts on the hyperparameter tuning using RoBERTa and XLSR-wav2vec 2.0 and the attention pooling. 

\begin{figure}[h]
     \centering
     \begin{subfigure}[b]{0.33\textwidth}
         \centering
         \includegraphics[width=\textwidth]{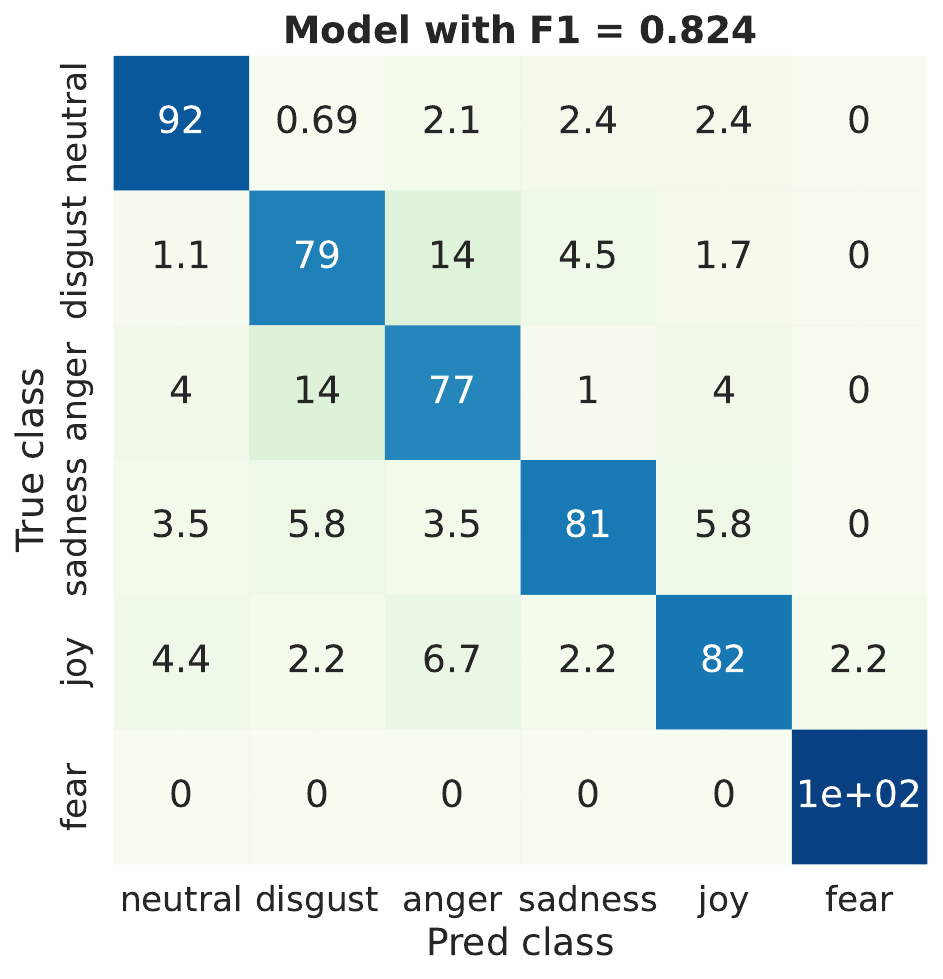}
         \caption{
            Hidden dense layers = 3\\
            Weight decay = 0.1}
         \label{fig:y equals x}
     \end{subfigure}
     \hfill
     \begin{subfigure}[b]{0.33\textwidth}
         \centering
         \includegraphics[width=\textwidth]{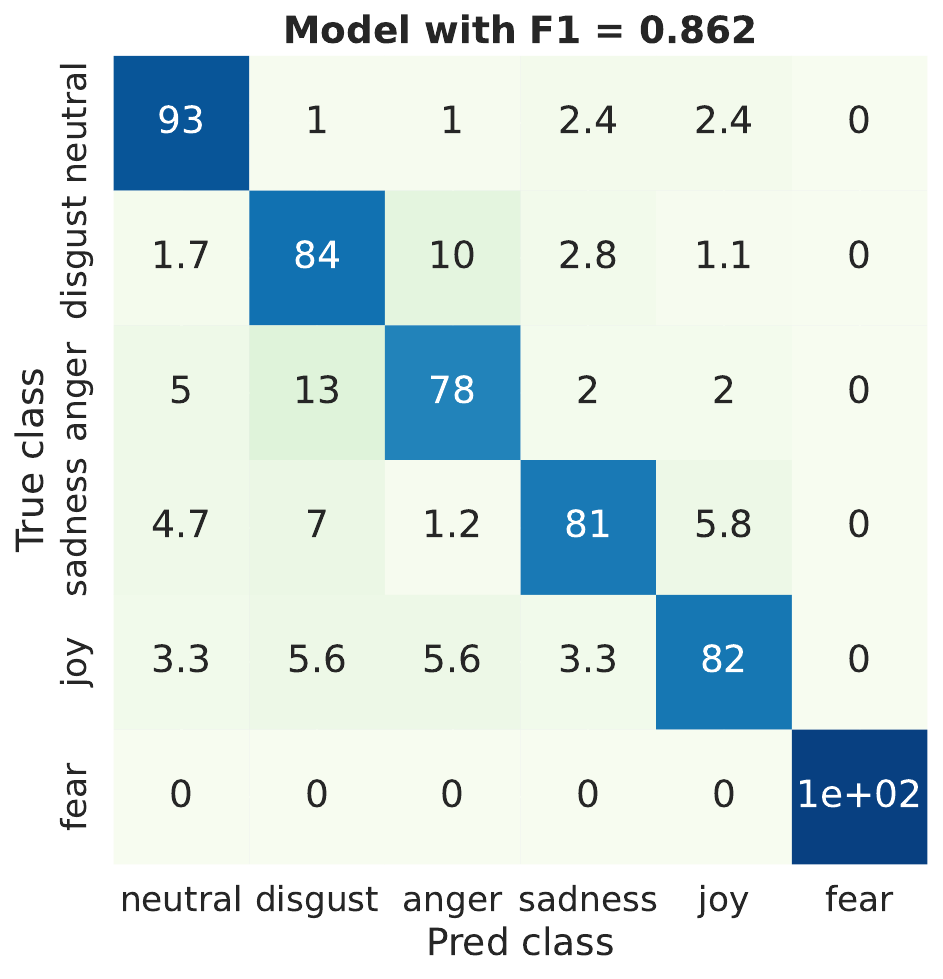}
         \caption{
            Hidden dense layers = 2\\
            Weight decay = 0.01}
         \label{fig:five over x}
     \end{subfigure}
     \hfill
     \begin{subfigure}[b]{0.33\textwidth}
         \centering
         \includegraphics[width=\textwidth]{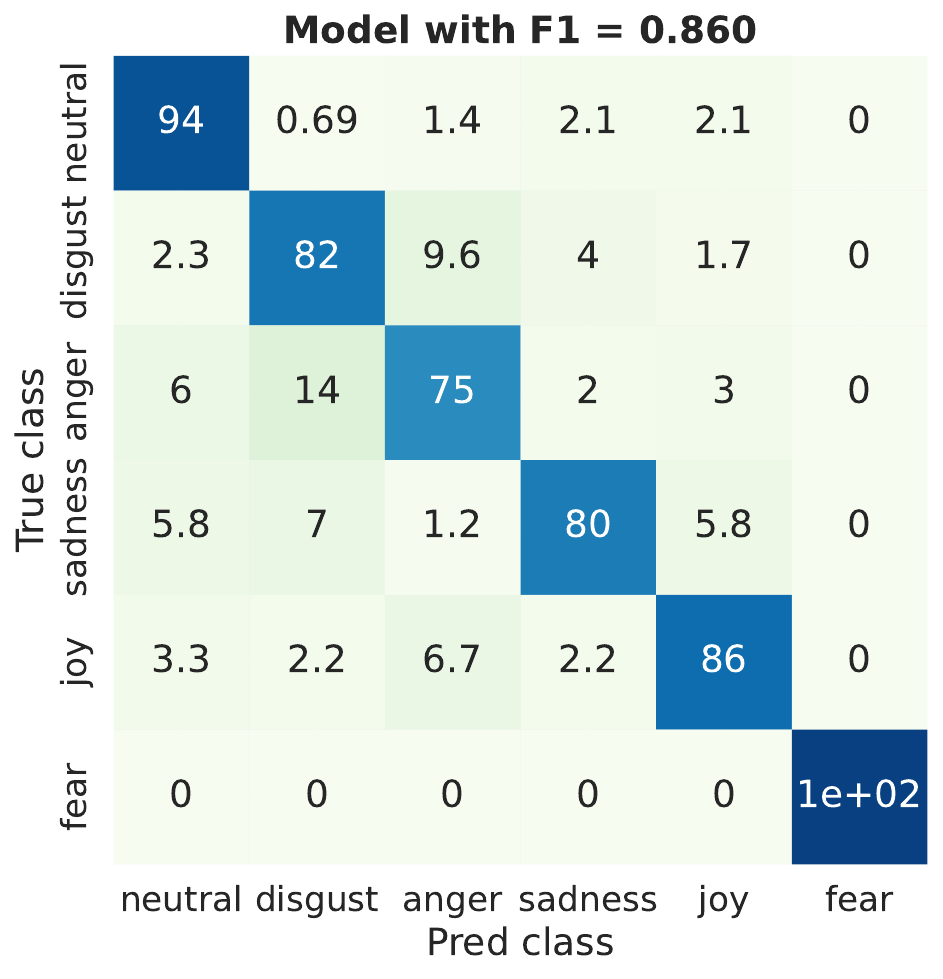}
         \caption{            
            Hidden dense layers = 2\\
            Weight decay = 0.1}
         \label{fig:three sin x}
     \end{subfigure}
        \caption{Confusion matrices in the test set. The drop-out was set to 0.1 and the data augmentation probability to 0.3.}
        \label{fig:confusion-matrices}
\end{figure}

The top three different models that obtained the best score achieved 86.20\%, 85.96\%, and 82.43\%. Their confusion matrices over the test set are displayed in Figure \ref{fig:confusion-matrices}. As can be seen, anger is the most difficult emotion to classify, normally getting confused with disgust. It is remarkable that, despite being very scarce in the dataset, fear is very separable from the rest of the emotional spectrum. 

Section \ref{sec:experimental-setup} remarked that thresholding techniques failed to outperform the models and were, therefore, discarded. Consequently, the only non-trainable approach used to improve the model's F1-Score was model ensembling. Initially, models with different feature extractors were employed to leverage the diversity of features and create a robust system. However, this approach proved to not be effective. Instead, the three best-performing models on the validation set were ensembled, improving the F1-score to 86.69\%. Table \ref{tab:model-ensemble} compares the results of these approaches with the challenge baseline.

\begin{table}[h]
\begin{tabular}{llll}
\toprule
\textbf{Model Name}     & \textbf{Hidden dense layers} & \textbf{Weight Decay} & \textbf{Test F1-Score} \\ \hline
Top 1 Model    & 2                            & 0.01                  & 86.20\%                  \\
Top 2 Model     & 2                            & 0.1                   & 85.96\%                  \\
Top 3 Model    & 3                            & 0.1                   & 82.43\%                  \\
Model Ensemble & -                            & -                     & 86.69\%                  \\
Baseline       & -                            & -                     & 53.08\%                  \\ \bottomrule
\end{tabular}
\caption{Results of the top 3 models, the model ensemble (hard voting), and the baseline of the challenge over the test set.}
\label{tab:model-ensemble}
\end{table}

\section{Conclusions and future work}
This study presented a multimodal model for the emotion recognition challenge EmoSPeech 2024 within the IberLEF 2024 framework, aimed at recognizing emotions from speech and text inputs. The architecture comprised two pre-trained models, one dedicated to speech and the other to text. They extracted feature vectors that the model concatenates into a unified hidden representation vector.
 On the one hand, for the audio side, different experiments were conducted with WavLM, XLSR-wav2vec 2.0, and HuBERT. On the other hand, the optimization of the textual component of the architecture involved exploration with the models BERT, XLM-RoBERTa for Spanish, BETO, and its finetuned version for emotion. The best performance is achieved by jointly combining RoBERTa and XLSR-wav2vec 2.0.

After the model concatenates the text and speech feature vectors, it employs a dimensionality reduction via reduced attention pooling. This mechanism, with fewer parameters than its standard counterpart,  facilitates the seamless integration of text and audio while mitigating the risk of overfitting to the training set. Subsequently, a stack of dense layers processed the output vector, using its compressed information to extract the class prediction. Additionally, to optimize performance and maximize the F1-Score in the competition, model ensembling techniques were adopted, employing hard voting on the top three models. In summary, the system was capable of achieving an F1-Score of 86.69\%, an absolute increment of 33.61\% compared to the baseline, and securing the first position in the challenge.

After the conclusion of this competition, continuing the research of new paradigms for Speech Emotion Recognition (SER) could be a captivating line of research. One of the lines is improving the efficiency of speech feature extractors. In the paper of WavLM, the authors claim that, in general, most self-supervised learning models (SSL) models focus primarily on Automatic Speech Recognition (ASR) tasks. However, by training an SSL model to jointly learn masked speech prediction and denoising in the pretraining stage, the model's capabilities extend beyond ASR, outperforming other SSL in fields such as SER.

This statement could seem contradictory because, in Section \ref{sec:results}, it is proven that XLSR-wav2vec 2.0 outperforms WavLM Large. Nevertheless, this outcome is likely due to  XLSR-wav2vec2.0 being trained with a very extensive multilingual dataset. If WavLM was trained with a comparable volume of data, it could potentially outperform XLSR-wav2vec 2.0, leveraging its joint learning of masked speech prediction and denoising to achieve superior performance in various tasks, including Speech Emotion Recognition.

\begin{acknowledgments}
This work has been promoted and financed by the Government of Catalonia through the Aina project, as well as by the Spanish \textit{Ministerio de Ciencia e Innovación} through the AdaVoice project (PID2019-107579RB-I00).

\end{acknowledgments}

\bibliography{bibliography}

\appendix

\end{document}